# Birth-Burst in Evolving Networks

**Dong Chen[1], Hong Yu[1,2,3]**

[1]University of Massachusetts, Amherst, MA

[2]University of Massachusetts Medical School, Worcester, MA

[3]Bedford VAMC, Bedford, MA

**Abstract**
The evolution of complex networks is governed by both growing rules and internal properties. Most evolving network models (e.g. preferential attachment) emphasize on the growing strategy, while neglecting the characteristics of individual nodes. In this study, we analyzed a widely studied network: the evolving protein-protein interaction (PPI) network. We discovered the critical contribution of individual nodes, occurring particularly at their birth. Specifically, a node is born with a fitness value - a measurement of its intrinsic significance. Upon the introduction of a node with a large fitness into the network, a corresponding high birth-degree is determined accordingly, leading to an abrupt increase of connectivity in the network. The degree fraction of these large (hub) nodes does not decay away with the network evolution, while keeping a constant influence over the lifetime. Here we developed the birth-burst model, an adaptation of the fitness model, to simulate degree-burst and phase-transition in the network evolution.

**Introduction**
Computationally modeling the evolution of complex networks has been an active research field since the Erdős-Rényi random graph model in 1959.[1] The early studies focused on how links are added to the networks. The construction of networks is either based on random linking of fixed nodes as in the Erdős-Rényi random graph model[1] or random rewiring of a circularly connected network as in the Watts-Strogatz small-world model.[2] The widely recognized Barabási-Albert (BA) scale-free model[3] has introduced a generative strategy to grow networks: (i) new nodes are generated sequentially, (ii) a constant amount of links are built between the new node and the existing nodes. The growth of the network follows the "preferential attachment" – the ability for an existing node to compete for connections depends on its degree, which results in a power-law degree distribution. Barabási and Albert have therefore introduced one of the most important features of real-world networks – the scale-free property that is evidenced prevalently in many real networks, including WWW,[4,5] Internet,[6] and citation networks.[7]

Preferential attachment, also known as cumulative advantage proposed earlier by Price in 1976,[8] leads to a simple result: "the rich get richer", resembling the Matthew effect in sociology. Because the nodes in the BA scale-free model are continuously added to the network, the early nodes has more opportunity to grow into large node, which is a phenomenon called "aging effect" or "first mover effect."[9,10] This effect was validated in citation networks where the first published paper has the advantage to attract more citations then a relevant paper published later.[10] However, the effect is not found in many real networks, where large degree nodes may continuously appear over the lifetime of network evolution.

In addition, it has been observed in realistic networks that the preferential attachment is suppressed by random connections between existing nodes, thus demonstrating a sublinear shape - in contrast to the linear shape of scale-free - in the "loglog plot" of degree distribution[11]. Subsequently many generalization models are proposed, including Krapivsky et al.,[12] Dorogovtsev and Mendes,[13] Krapivsky and Redner.[14] Most of the models, however, consider only the adjustment of node degree, ignoring the other properties that may contribute to the network evolution.

The fitness model, adopted the formulation of Bose-Einstein condensation, was proposed by Bianconi and Barabási[15] to address the significance of the single node. Unlike the scale-free model in which nodes are identical, each node in the fitness model is assigned a weight named "fitness," reflecting its significance. In the fitness model, the connection probability of an existing node is dependent on both the degree, as in the scale-free model, and its fitness value. Such a model is able to overcome the limitations of "aging effect" so that later joined nodes with a large fitness have opportunities to transcend into large degree nodes.

Depending on the fitness distribution, networks may present various phases. A uniformly distributed fitness degenerates into scale-free. Otherwise, the nodes with large fitness have the chance to get richer and therefore the network presents a fit-get-rich phase. A node with a large fitness may take a significant fraction of the total degrees of the network. The largest node is an absolute winner in contrast to the other nodes. The network therefore presents a winner-takes-all phase that was illustrated by the Bose-Einstein condensation model.[15] The concept of the phases in complex networks including the scale-free as a special case is particularly suitable in describing the evolving process of the growing networks.[16,17]

In this paper, we studied an important and widely studied biomedical network: the protein-protein interaction (PPI) networks[18-24]. We analyzed how the PPI network evolves, focusing on the contribution of individual nodes. We observe that the deviation from the scale-free property cannot be explained with the existing models. For example, we observe that some nodes are born with large degrees and play a critical role in the lifetime of its evolution. We here introduce the 'birth-burst' model to fit the PPI network evolution.

**Results**
**Protein-protein interaction network evolution**
The interactions between proteins are critical in many biological processes – molecule synthesizing, gene expression, metabolic pathways, etc. The mapping and understanding of protein-protein interaction networks facilitate a systematical representation and interpretation of these processes.25,26 Although large amount of protein interactions in many species are continuously discovered, many concerns and debates are still not properly addressed. The completeness of the PPI is one of the most discussed. Correctly formulating of the evolution rules in these networks in the key to predict their continuous construction. In this study, we adopted PPI networks from the expert-annotated database BioGRID27 that curated both genetic and physical interactions. BioGRID comprises PPI networks from several species, including Saccharomyces Cerevisiae PPI network (SC-Net),24 one of the most studied and high-quality PPI networks and Homo Sapiens PPI network (HS-Net), a fast growing PPI network.

As shown in Fig. 1, the degree distributions of both networks follow an overall power law in the early stage of development, followed by an apparent deviation before the networks evolve into a mature stage. The sublinear distribution on the "loglog" plot indicates a highly suppressed preferential attachment induced by the large amount of random connections between existing nodes, which is evidenced in by tracking the continuous degree construction. As the power-law degree distribution is one of the most important topological feature of the scale-free network,28,29 the PPI networks thus can not be well described by the BA scale-free model. Another phenomenon worth of mentioning is that there are large numbers of nodes with small degree, which disobeys the minimum birth degree setup in the BA scale-free model (Fig. 2C). The fitness model30 (Fig. 2D) that shows a superlinear degree distribution can neither explain the topological characters in the PPI networks.

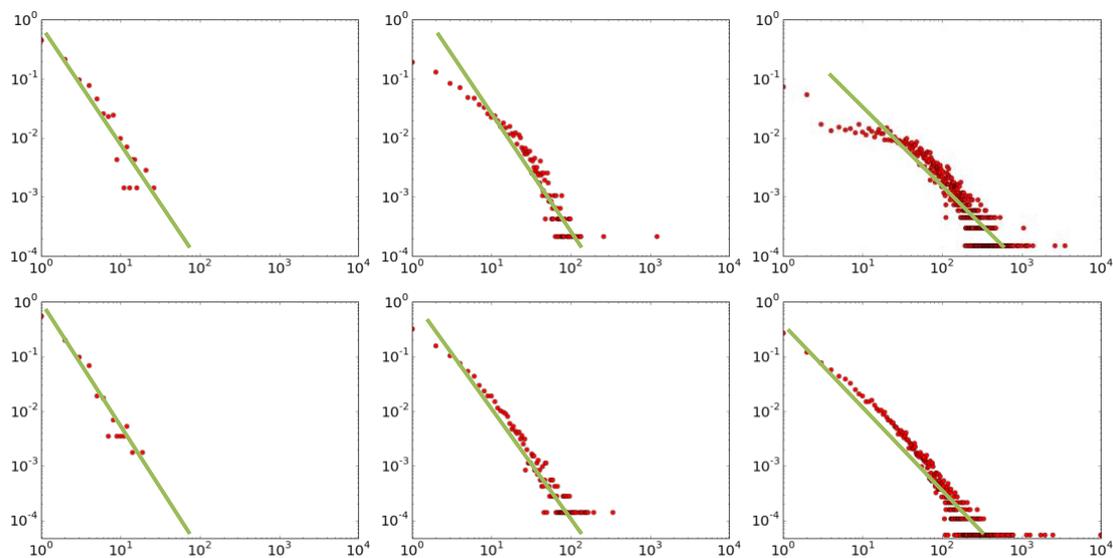

**Figure 1.** The degree distribution of Saccharomyces Cerevisiae (upper sequence) and Homo Sapiens (lower sequence) protein-protein interaction (PPI) networks evolve over time: 1993 (left), 2003 (middle), and 2013 (right).

To reveal the construction of the PPI networks, we tracked the degree growth of the individual nodes in these networks. One of the most important features discovered during the network evolution is that some newly introduced nodes suddenly brought a large number of connections to the system, resulting an abrupt degree increase. Following the degree burst, their degree fractions of the entire system gradually decay to a nonzero plateau, a phenomenon described by

Bose-Einstein condensation[15]. In Fig. 3, we plotted the progress of degree fraction variation of top four hub nodes in both Saccharomyces Cerevisiae and Homo Sapiens PPI networks, which represented the birth events that is prevalent in the PPI evolution. The large birth degree in PPI networks can be treated as a novel event in the process of network growth, in which, a particular protein becomes particularly interested due to its functions, large amount of interactions to the other proteins were discovered in a short period of time. After that, the connections of this node were still growing, but the growth rate gradually reduces to a steady level. Due to the significance (large fitness) of this node, it still attracts a consistent attention over its lifetime. The largest node occupies a finite fraction of degree of the entire system is modeled by Bose-Einstein condensation in the fitness model.[15] In PPI network, it is not only the largest nodes; many hub nodes occupy a finite fraction of degrees long after their birth burst (Fig. 3).

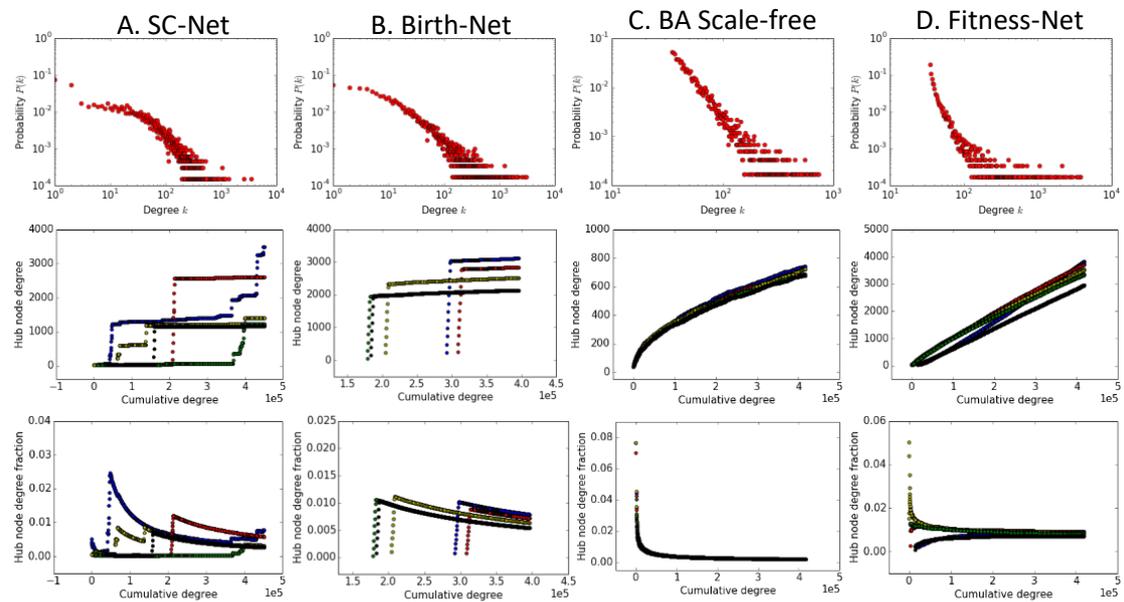

**Figure 2**. Degree distribution (top), hub node degree growth (middle) and hub node degree fraction variation (bottom) of various networks: (A) realistic networks, (B) birth-net and two existing models – (C) BA scale-free and (D) fitness networks. Hub nodes in SC-net presented apparent degree burst followed by preferential attachment. The degree fraction grows into a stable stage after the burst. The preferential attachment is greatly suppressed capering to the scale-free and fitness model. The birth model simulates the degree burst and slow growth rate while maintains a stable growth state as in the fitness model.

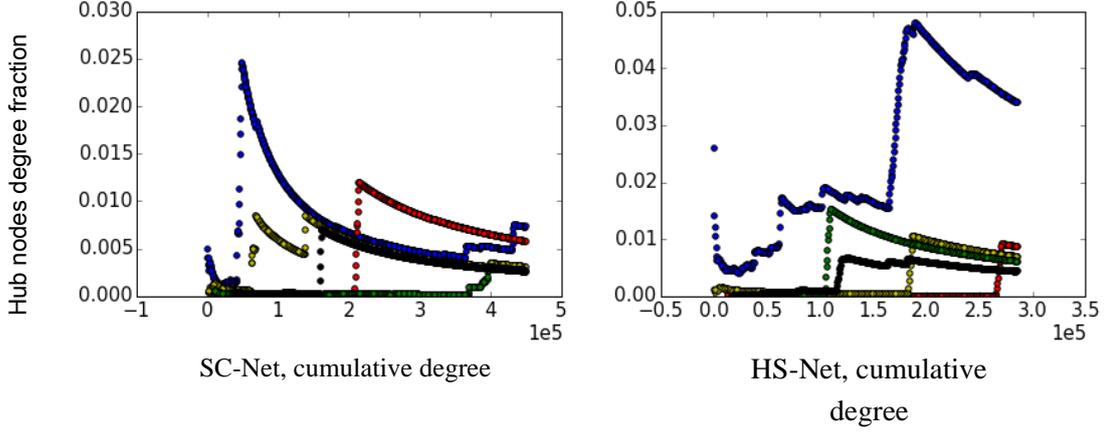

**Figure 3.** Hub nodes degree burst in the process of evolution of Saccharomyces Cerevisiae (left) and Homo Sapiens (right) networks. At some point, a large node suddenly brings a large amount of degrees. Before this point, the node has nearly zero degree. After this point, the large node degree ratio gradually decreases to the stable stage, occupying a finite fraction of total degrees.

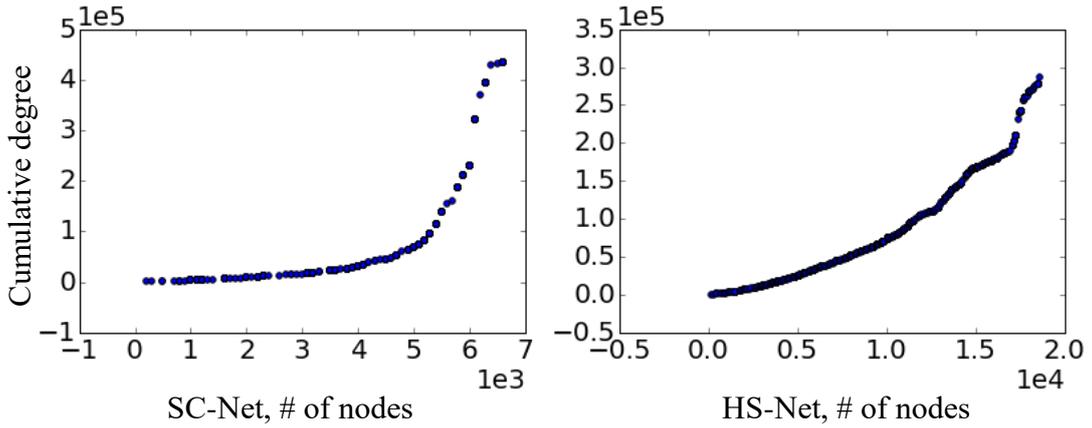

**Figure 4.** Cumulative degree of the network increases faster than linear with respect to the node expansion.

Nodes in the BA scale-free model are generated with a constant birth degree, which resulted in a linear growth of the cumulative degree of the system. In realistic networks, such as the PPI networks, there is a varied birth degree associate with each node. The growth rate of the degree with respect to the nodes in PPI networks indicates the activity of the entire system. With the development of the experimental methodology, the interactions between proteins are discovered faster than before. The overall connection and birth degree would increase accordingly. Thus we formulate the network activity into the average degree per node, denoted as $d$ that is dependent on the time or the number of nodes in the system. In the PPI networks, the growth of the cumulative degree is well described in an exponential formula as in equation (1).

$$d = c\, n^{\beta} \tag{1}$$

Where c is a constant, $\beta$ reflects the activity. $\beta$ decays to zero, as the degree growth rate is linear.

**Birth-burst model**

The birth degree reflects the initial attraction of a new node, which closely related to its intrinsic property. To emphasize the significance of the node, a value of fitness $\eta$ is assigned to each node. By introducing the fitness, each node differentiates from the others not only by their age, but also by their internal strength. A large fitness induces a large birth degree, further enhancing the ability to build connections. Both the birth degree and the large fitness provide the opportunity of the later joined node to overcome the first-mover advantage of the early nodes.

To address the significance of the nodes and the birth degree **m**, we developed the birth-burst model. We consider node generation as a birth event. Other than a constant birth degree, each node has its own birth degree according to its fitness. A simple linear relationship of the birth degree and the fitness is empirically determined as in equation (2).

$$m_i = d\,\eta_i$$

(2)

The birth-burst model is a natural extension of the fitness model. The probability of an old node having one of the connections of the new node is determined by equation 3. In addition to the linear relationship with its degree, the connection probability of an existing node is dependent on its fitness.

$$\Pi_i = \frac{\eta_i k_i}{\sum \eta_i k_i}$$

(3)

Where $\Pi_i$ is the connection probability of an existing node, and $\eta_i$ and $k_i$ are its corresponding fitness and degree.

**Discussion and conclusions**

The fitness distribution can be measured with the same method as the measurement of preferential attachment.[11] For simplicity, the equation 4 is used as the fitness distribution, where $\gamma$ controls the relative amount of large-fitness nodes. A large $\gamma$ suppresses the amount of node with large fitness. It is easy to prove that when the $\gamma$ greater than 1, the system falls into the fit-takes-all phase of Bose-Einstein condensation. The largest node occupies a finite fraction of degrees. In fact, many hub nodes with large fitness fall into this category. They maintain a stable growth after the birth burst.

$$f(\eta) = (1+\gamma)(1-\eta)^\gamma$$

(4)

The fitness of the protein in the PPI networks can be interpreted in several aspects. The importance of the protein itself is an obvious measurement. Important proteins may interact with many other proteins to perform physiological functions. Since all protein-protein interactions are presented by the publications, the discovery of important interactions may attract attention of scientists, who further study the corresponding proteins, contributing to short time degree-burst. Thus the fitness and the birth degree are also related to the public attentions of research.

Following the BA scale-free model, one can create an evolving network that follows the power-law distribution. The BA scale-free model has been successfully used to explain the scale-free property of many networks. However, the generation of realistic networks has many characteristics that cannot be revealed by the BA scale-free model. For instance, the PPI network (Fig 1) has a degree distribution that is more scattered, deviated from the power law. Although the fitness model (Bose Einstein model) put more emphasis on the nodes themselves, it can neither explain the deviation from power-law in general (Fig 2D). The nonlinearity of degree distribution is highly affected by the random connections between existing nodes in addition to the preferential attachment. As shown in the real PPI networks, the preferential attachment is highly suppressed; the degree distribution presented a sublinear shape, which is due to the large amount of random connections between existed nodes. This resulted in a much slower growth rate of the node degree after birth comparing to the BA model and fitness model (Fig 2). The birth event of each node is more critical in determining the PPI network evolution. It's also found that the hub nodes always developed from the early stage in both Barabasi-Albert scale-free model and fitness model. This is a 'first-mover' or 'First-Get-Richer' effect, mainly caused by the property of preferential attachment. Even the later joined nodes have large fitness; they have little chance to beat the early nodes that have already built large number of degrees. However, it is not the case in realistic networks, the hub nodes happen randomly in all stages of network evolution. A large birth degree may overcome the long-time accumulation of early nodes. In order for this to happen, the preferential attachment must be greatly suppressed as in equation 5. $\alpha$ ($\leq 1$) reflects the amount of random connections in the networks. Smaller number of $\alpha$ indicates a high randomness and low preferential attachment in the system.

$$\Pi_i = \frac{(\eta_i k_i)^\alpha}{\sum (\eta_i k_i)^\alpha} \qquad (5)$$

Neither the fitness model nor the scale-free model describes well the internal connections (the connections between existing nodes) other than the connections bring purely by the newly introduced nodes. The internal connections are the dynamics for most active networks. In the process of protein-protein network construction, new interactions are consistently discovered between two proteins that are already in the network. In WWW networks, existing webpages connect to another existing webpages. An existing node may search for the highly popular nodes (high degree $k$) or highly quality nodes (high fitness $\eta$). However, these connections often present a random style or highly suppressed preferential attachment. The distribution of these networks presents a sublinear shape in the "loglog plot" of degree distribution that is found consistently in realistic networks. In addition, the preferential attachment of new nodes is also high suppressed if we measure it in the realistic evolving networks. The suppressed preferential attachment can be simply modeled by a slight modification to the fitness model. Instead of a linear dependence of the attachment probability to the fitness and degree, the attachment probability is decreased by the exponent $α$ ($α < 1$) that controls the randomness of the connection construction while still maintain the sequential addition of nodes and connections.

The birth-bust model addressed the varied "birth degree" that is more realistic – an attractive node (large fitness) has the potential to build more connections start from birth. It is more critical in the PPI networks, where a large amount of connections happens at birth time. Our model can explain this "burst event" by simply introducing the birth degree. The general degree distribution and evolution of the PPI networks are thus recovered accurately (Fig. 2B Birth-Net).

It's worth of mentioning that in the PPI networks, the degree burst does not happened at birth time, but latter in any time of the network evolution. The early development of these hub nodes is subtle. To make the simulation procedure clear and neat, the early degree growth are cumulated to the degree burst time. We treat the degree burst time as the birth or rebirth time, so that each node is born with a birth degree, followed by a suppressed preferential growth.

Some other models have been proposed to explain the differences between biological networks and non-biological networks. In the node duplication model[31], the scale-free or power-law property is predetermined and the deviation of exponents are explained. However, the analysis of the PPI networks presented in shows dramatics different degree distraction that can be well explained by the birth-burst model.

**Future Work**

We speculate that our birth-burst models are robust and can apply to other networks, including citation, cooperation and other social networks, and we plan to validate our hypothesis in the future work. We also speculate that the degree burst, hub node location and random connections can be predicted according to the node property and the current network structure and its evolution. We plan to develop such a simulation model in our future work.

**Methods**

**Fitness distribution measurement**
Following equations (3), one can measure the fitness distribution with a similar method described by Jeong et. al. [11] Firstly, we count the degree of the existing nodes as the network has grown into a mature stage (i.e. 2012 in SC PPI network). Secondly, we count new connections in a short period (i.e. 2013 in the SC PPI network). Since the probability of new connection is proportional to the fitness as well as the existing degree, we can simply use the new connections divided by the corresponding existing degree to acquire the fitness. Note that what we get is the relative amount of fitness, but it's enough to achieve the fitness distribution. For simplicity we could use a similar distribution as in equation 4, where $γ$ controls the distribution of large and small nodes. If there are large amount of small nodes, $γ$ is large enough to guarantee a substantial amount of node with less degrees. It is easy to prove that when the $γ$ greater than 1, the system falls into the "winner-takes-all" phase of Bose-Einstein condensation[15]. The largest node occupies a finite fraction of degrees. In fact, many hub nodes with large fitness fall into this category (Fig. 2D). They maintain a stable growth for the lifetime after the birth burst.

**Birth-burst network construction**

The construction of the birth-burst network relies on the information from the existing networks. A fitness distribution is firstly acquired from the network following the procedure described above. The average degree is measured by counting the existing nodes and the entire degree of the networks. The network generation started by the assignment of the fitness to a newly introduced nodes with a corresponding birth degree. The connection probability of the existing nodes follows the suppressed preferential attachment as in equation 5. The value of $α$ is determined by the random connections between the existing nodes, which can be empirically learned by the representation of the hub nodes evolution.

**Acknowledgement**

This work is supported by National Institutes of Health (http://www.nih.gov/) with the grant number 1R01GM095476. The funders had no role in study design, data collection and analysis, decision to publish, or preparation of the manuscript. We thank Cris Moore for his comments.

## Additional information

### Author contributions

D.C. and H.Y. wrote the main manuscript text and D.C. prepared the figures. All authors reviewed the manuscript.

**Competing financial interests**

The authors declare no competing financial interests.